\documentclass[pra,twocolumn,aps,superscriptaddress,showpacs]{revtex4-2}
\usepackage{hyperref}
\usepackage{graphicx}
\usepackage{amsmath}
\usepackage{amsfonts}
\usepackage{amssymb}
\usepackage{epsfig}
\usepackage{subfigure}
\usepackage{mathtools}
\usepackage{braket}
\usepackage[usenames,dvipsnames]{color}
\usepackage{setspace}
\usepackage{lineno}
\usepackage{times}
\hypersetup{
      colorlinks=true,
      citecolor=blue,
      linkcolor=blue,
      urlcolor=blue}

\usepackage{filemod}

\def\be{\begin{equation}}
\def\ee{\end{equation}}
\def\bg{\begin{equation}\begin{gathered}}
\def\eg{\end{gathered}\end{equation}}

\begin{document}
\title{Staggered quantum phases of dipolar bosons at finite temperatures}
\author{Kuldeep Suthar}
\email{Corresponding author. kuldeep@gate.sinica.edu.tw}
\affiliation{Institute of Atomic and Molecular Sciences, Academia Sinica, 10617 Taipei, Taiwan}
\author{Kwai-Kong Ng}
\email{Corresponding author. kkng@thu.edu.tw}
\affiliation{Department of Applied Physics, Tunghai University, Taichung 40704, Taiwan}

\date{\today} 

\begin{abstract}
  The extended Bose-Hubbard model with correlated tunneling exhibits staggered 
superfluid and supersolid quantum phases. We study finite-temperature phase 
transitions of quantum phases of dipolar bosons in a two-dimensional optical 
lattice using Gutzwiller mean-field and quantum Monte Carlo approaches. When 
nearest-neighbor repulsion is comparable to the on-site interaction, we find 
that the two topologically distinct superfluids are separated by a normal fluid 
phase, while at stronger off-site interactions, density-modulated insulating 
quantum phases appear. We estimate the critical temperature of the staggered 
superfluid to normal fluid transition and show that this transition is of 
the Kosterlitz-Thouless type. Finally, we elucidate the coexistence of 
staggered quantum phases in the presence of an external trapping potential. Our 
study paves a way to observe novel staggered quantum phases in recent dipolar 
optical lattice experiments.
\end{abstract}

\maketitle

\section{Introduction}
  The ultracold atomic gases trapped in an optical lattice provide a powerful 
tool to simulate the low-energy behavior of the effective Hamiltonians of 
condensed matter models~\cite{lewenstein_07,bloch_08,bloch_12,gross_17}. 
Excellent experimental control over the model parameters led to the observation
of the Mott insulator (MI) to superfluid (SF) phase 
transition~\cite{greiner_02}, realization of the Hofstadter 
model~\cite{aidelsburger_13}, and various novel phenomena in many-body 
physics~\cite{schafer_20}. This set-up can also be used to investigate effects 
that are not possible in conventional solid-state physics, such as tuning the 
interparticle interaction strength~\cite{chin_10} and generating very strong 
effective magnetic fields~\cite{aidelsburger_18}. The Bose-Hubbard model (BHM) 
describes interatomic interactions at a lattice site. However, other 
interaction processes also affect the properties of strongly-correlated 
materials. In particular, the bond-charge interaction of the extended Hubbard 
model~\cite{luz_96} has been invoked to explain various phenomena including 
high-temperature superconductivity~\cite{hirsch_89,essler_92,appel_93}. Due to 
the lack of precise control over the interaction strengths, as well as the 
complexity of materials, interaction-induced phenomena cannot be probed in 
condensed matter physics. 

 The unique features of the ultracold atoms and optical lattices create an 
ideal platform to study complex phenomena due to interparticle particle 
interactions~\cite{bloch_05,gross_17}. For long-range interacting atomic gas, 
the introduction of nearest-neighbor (NN) interaction induces charge 
density-wave (CDW) and supersolid (SS) ground states, which spontaneously break 
the translational symmetry of the lattice~\cite{lahaye_09,baranov_12}. At 
higher average atomic densities, the onsite interaction results in higher 
order tunneling processes. One such effect is density-induced tunneling 
(DIT) which is analogous to the bond-charge interaction of fermions. This 
considerably affects the properties of soft-core dipolar bosons. Theoretical 
investigations have shown the influence of DIT on MI-SF quantum phase 
transitions~\cite{mark_11,luhmann_12,pilati_12}, Bose-Fermi multicomponent 
mixtures~\cite{ospelkaus_06,gunter_06,best_09,jurgensen_12,luhmann_15}, band 
structures~\cite{luhmann_12}, nonequilibrium dynamics~\cite{jurgensen_14}, and 
the emergence of staggered quantum 
phases~\cite{luhmann_16,johnstone_19,kraus_20,suthar_20}. The density-induced 
tunneling has been observed in recent quantum gas 
experiments~\cite{jurgensen_14,baier_16}. The ramp of the lattice potential and 
the heating mechanism during time-of-flight measurements contribute to thermal 
fluctuations in experiments. The excited states in the system affect the 
parameter region of quantum phases and a normal fluid (NF) state appears in 
the system at finite temperatures. Several previous studies have discussed the 
role of finite temperature on the homogeneous system for MI-SF transition of 
BHM~\cite{spielman_07,sansone_08,rigol_09,garcia_10,mahmud_11,sajna_15,
pires_17} and CDW-SS transition of extended 
BHM~\cite{suthar_20a,bai_20,chen_20}. Moreover, the effects of the trapping 
potential on the coexistence of uniform and density-modulated phases are also 
examined for short-~\cite{mahmud_11,ceccarelli_12,gupta_13} and long-range 
many-body systems~\cite{landig_16,suthar_20a,lagoin_22}. Since experiments are 
performed at finite temperatures, it is imperative to consider the role of 
thermal fluctuations in determining the thermal and insulating 
regions~\cite{parny_12,parny_17}. Moreover, the effects of finite temperatures 
on the staggered superfluidity of strongly-correlated quantum many-body systems 
remain unanswered.  

 In the present work, we investigate the finite-temperature phase diagram of 
a homogeneous two-dimensional extended Bose-Hubbard model with correlated 
hopping. In particular, we focus on higher average number densities where  
interaction-induced processes lead to staggered quantum phases. Our results 
show the role of thermal fluctuations resulting in a NF state which intervenes 
in two different kinds of superfluidity, one due to single-particle hopping 
while the other is due to DIT. In order to have experimental relevance, we 
examine both the finite-size effects and the phase coexistence caused by the 
trapping potential. The thermal phase transition of staggered superfluidity 
follows the Kosterlitz-Thouless (KT) transition. We discuss the combined 
effects of trapping and finite-temperature and identify the parameter regime 
of staggered superfluidity. 

 The paper is organized as follows. In Sec.~\ref{ham_methods} we introduce the 
model Hamiltonian of isotropic dipolar bosons and describe the approaches used 
in the present work. We discuss the finite-temperature phase diagrams for weak 
and strong off-site interactions in the presence of correlated tunneling in 
Sec.~\ref{pd_den}. In Sec.~\ref{fss_scal} we further examine the finite-size 
effects of staggered quantum phases using the quantum Monte Carlo (QMC) 
approach. The coexistence of staggered quantum phases at finite temperatures in 
an external harmonic trapping potential is discussed in Sec.~\ref{trap}. 
Finally, we conclude in Sec.~\ref{conc}. 

\section{Hamiltonian and methods}
\label{ham_methods}
\subsection{Extended Bose-Hubbard model with correlated hopping}
\label{eBHM}
We consider spinless bosons confined to a two-dimensional square optical 
lattice. The atoms can hop between nearest-neighbor sites of the lattice and 
experience on-site repulsion. The dipolar interaction considered here is 
isotropic in nature and as a minimal model, the atoms of nearest-neighboring 
sites interact repulsively. The model Hamiltonian reads as~\cite{baier_16}  
\begin{eqnarray}
  \hat{H} &= &-t\sum_{\langle i,j\rangle}\left(\hat{b}^{\dagger}_{i}\hat{b}_{j} 
	      + \rm{H.c.} \right) 
	  + \frac{U}{2}\sum_{i}~\hat{n}_{i}~(\hat{n}_{i} - 1) \nonumber \\
        &&+ \sum_{\langle i,j\rangle} \bigg[V~\hat{n}_{i}\hat{n}_{j} 
	  - t'~\hat{b}^{\dagger}_{i}(\hat{n}_{i}+\hat{n}_{j})\hat{b}_{j}\bigg] 
	  - \mu \sum_{i}~\hat{n}_{i}.  
\label{ebhm}
\end{eqnarray}
The bosonic operator $\hat{b}^{\dagger}_{i}$ ($\hat{b}_{i}$) creates 
(annihilates) an atom at $i$th lattice site, and 
$\hat{n}_{i} = \hat{b}^{\dagger}_{i} \hat{b}_{i}$ is the corresponding number 
operator. The first term describes the kinetic energy with $t$ as the hopping 
strength between nearest-neighbor sites $i$ and $j$ on a square lattice with 
periodic boundary conditions. The second term represents the on-site repulsive 
interaction between atoms with strength $U$. The third term is the dipolar  
interaction which includes the nearest-neighbor repulsive interaction $V$ and 
density-induced tunneling with strength $t'$, depending on the density of atoms 
at each site. The last term $\mu$ denotes the chemical potential which controls 
the atomic density in the grand canonical ensemble.

\subsection{Methods}
\label{methods}
To study the properties of the system at finite temperatures, we first use the 
Gutzwiller mean-field 
approach~\cite{gutzwiller_63,rokshar_91,krauth_92,sheshadri_93,bai_18}. In the 
mean-field approximation, the bosonic annihilation operator is decomposed as 
$\hat{b}_{i} = \langle \hat{b}_{i} \rangle + \delta \hat{b}_{i}$ where 
$\langle \hat{b}_{i} \rangle \equiv \phi_{i}$ is the mean-field, also referred 
to as the superfluid order parameter, and $\delta \hat{b}_{i}$ is the 
fluctuation operator. A similar decomposition for the creation operator can be 
defined. Using this approximation, the Hamiltonian decouples the sites and all 
the off-site contributions are incorporated through the mean-field. The 
many-body Gutzwiller wave function is 
\begin{equation}
 |\Psi_{\rm GW}\rangle = \prod_{i}|\psi_{i}\rangle =\prod_{i}
		       \sum_{n}^{n_{\rm max}} c^{i}_n|n_i\rangle,
 \label{gw}
\end{equation}	
where ${|n_i\rangle}$ is occupation basis state with $n$ atoms at $i$th site, 
and we introduce a cut-off $n_{\rm max}$ on the maximum number of bosons per 
site, and $c^{i}_n$ are the complex coefficients for the state 
$|\psi_{i}\rangle$. The $|\Psi_{\rm GW}\rangle$ is normalized by the 
corresponding complex coefficients $c^{i}_n$. At zero temperature, the SF order 
parameter is a measure of the off-diagonal long-range order or long-range phase 
coherence $\phi_{i} \equiv 
\langle\Psi_{\rm GW}|\hat{b}_{i}|\Psi_{\rm GW}\rangle 
= \sum_{n} \sqrt{n} {c^{*i}_{n-1}}c_n^{i}$. It is finite for the SF phase while 
zero for the incompressible phase. The average density $\rho$ is given by 
$\sum_{i} n_{i}/L^2$ where $L$ is the system size. The atomic density at $i$th 
lattice site is $n_{i} \equiv 
\langle\Psi_{\rm GW}| \hat{n}_{i} |\Psi_{\rm GW}\rangle 
= \sum_{n} n |c_n^{i}|^2$. The density-assisted correlation or DIT order 
parameter is 
\begin{eqnarray}
 \eta_{i} &=& \langle\Psi_{\rm GW}|\hat{n}_{i}\hat{b}_{i} |\Psi_{\rm GW}\rangle
	  = \sum_{n} \sqrt{n} (n-1) {c^{*i}_{n-1}}c_n^{i},
\end{eqnarray}
which is finite for phases with finite SF order parameter. 

At finite temperatures, the presence of thermal fluctuations significantly 
modifies the phase transitions of the 
system~\cite{krutitsky_16,pal_19,suthar_20a,suthar_21}. The order parameters 
defining the superfluidity and density-dependent transport properties are given 
by their thermal averages. We retain the entire energy spectrum $E^{l}_{i}$ and 
eigenstates $\ket{\psi}^{l}_{i}$ obtained from the diagonalization of 
mean-field Hamiltonian. Then, the partition function of the system is 
$Z = \sum_{l} e^{-\beta E^{l}}$, where $\beta = (k_{B}T)^{-1}$ is the inverse 
of thermal energy at temperature $T$. The thermal average of the order 
parameters is
\begin{subequations}
  \begin{eqnarray}
    \langle \phi_{i}\rangle = \frac{1}{Z}\sum_{l}
                              \prescript{l}{i}{\bra{\psi}}
			      \hat{b}_{i} e^{-\beta E^{l}} \ket{\psi}^{l}_{i},\\
    \langle \eta_{i}\rangle = \frac{1}{Z}\sum_{l}
                              \prescript{l}{i}{\bra{\psi}}
			      \hat{n}_{i}\hat{b}_{i} 
			      e^{-\beta E^{l}} \ket{\psi}^{l}_{i},
  \end{eqnarray}
\end{subequations}
where $\langle \cdots \rangle$ represents the thermal averaging of the 
observable. Additionally, the average local occupancy at finite $T$ is defined 
as 
\begin{equation}
  \langle \hat{n}_{i} \rangle = \frac{1}{Z}\sum_{l}
                                \prescript{l}{i}{\bra{\psi}}
	  		        \hat{n}_{i} e^{-\beta E^{l}}\ket{\psi}^{l}_{i},
\end{equation}
from which the average density at finite $T$ is 
$\rho = \sum_{i}\langle \hat{n}_{i} \rangle/L^2$. The introduction of finite 
$T$ leads to the appearance of the NF phase. The NF phase is a compressible 
phase and it can be distinguished from other insulating MI and CDW phases by 
inspecting the compressibility~\cite{mahmud_11,parny_12}. The compressibility, 
which is the measure of local density variance, is 
$\kappa = \sum_{i} \left(\langle \hat{n}^2_{i} \rangle 
           - {\langle \hat{n}_{i} \rangle}^2 \right)/L^2$.
At finite $T$, due to the presence of thermal fluctuations, the insulating 
phases appear with nearly integer site occupancy whereas NF phase has real 
occupancy. Ideally, $\kappa$ is zero for insulating MI and CDW phases while 
it is nonzero for the NF phase. In the mean-field approach, the crossover 
between the insulating MI or CDW phase and NF phase is determined by the value 
of $\kappa$. Here, we identify NF phase by 
$\kappa > 10^{-5}$~\cite{buonsante_04,mahmud_11} and $\phi=\eta=0$. The zero SF
order parameter also distinguishes NF phase from other compressible phases. 
Furthermore the single-particle correlation and structure factor which are the 
measure of off-diagonal and diagonal long-range order, for momentum wave-vector
$\textbf{k}$ are 
\begin{subequations}
  \begin{eqnarray} 
    M (\textbf{k}) &=& \frac{1}{L^2} \sum_{j,j'} 
		       e^{i \textbf{k}\cdot(\textbf{r}_ j - \textbf{r}_{j'})}
		       \langle \hat{b}^{\dagger}_{j} \hat{b}_{j'} \rangle, \\
    S (\textbf{k}) &=& \frac{1}{L^2} \sum_{j,j'} 
        	       e^{i \textbf{k}\cdot(\textbf{r}_j - \textbf{r}_{j'})}
		       \langle \hat{n}_{j} \hat{n}_{j'} \rangle.
\end{eqnarray} 
\end{subequations}

Complementary to the Gutzwiller mean-field approach, we also employ the quantum 
Monte Carlo method to study the finite temperature phase diagrams of the 
model. In principle, at finite $T$, QMC is numerically exact and only subjects 
to the finite-size effect and statistical error if the infamous sign problem is 
absent~\cite{Pollet_2012}. Unfortunately, in our extended BHM $\hat{H}$, a 
negative value of the DIT ($t'$) competes with the single-particle hopping 
$t$ ($t>0$) that the overall effective hopping matrix elements,  
\begin{eqnarray}
  \langle n_i+1, n_j-1 | - t  \hat{b}^\dagger_i \hat{b}_j 
  - t'\hat{ b}^\dagger_i(\hat{n}_i+\hat{n}_j) \hat{b}_j |n_i, n_j \rangle \\ 
  \nonumber
  = [-t-t'(n_i+n_j-1)]\sqrt{n_i+1}\sqrt{n_j} ,
\label{effhop}
\end{eqnarray}
can be either positive or negative, depends on the neighboring local particle 
densities $n_i$ and $n_j$. This leads to possible opposite signs in the QMC 
simulation that cannot be removed with some simple transformation. In that 
case, the statistical uncertainty tends to diverge at low enough temperatures. 
Nevertheless, for most of the parameter regime and temperature range we 
investigate in this work, the average sign is not close to zero, therefore the 
sign problem is manageable to provide reliable observable measurements. 

We adopt the well-established stochastic series expansion approach with a
directed loop algorithm to simulate the model Hamiltonian 
$\hat{H}$~\cite{PhysRevB.59.R14157,PhysRevE.66.046701}. As usual, the 
superfluidity is measured via the fluctuation of the winding numbers $W_x$ and 
$W_y$. This is given by the number of times the off-diagonal operator list is 
wound around the 
lattice~\cite{PhysRevB.56.11678,PhysRevB.36.8343,PhysRevB.39.2084}
\begin{equation}
  \rho_s =\frac{1}{2\beta} \left(\langle W^2_x \rangle 
	                       + \langle W^2_y \rangle\right).
\end{equation}
On the other hand, the computation of the off-diagonal order parameter 
$M(\bf{k})$ requires the measurement of the matrix elements 
$\langle \hat{b}^{\dagger}_{j} \hat{b}_{j'} \rangle$ which can be carried out 
directly during the loop update~\cite{PhysRevE.64.066701}. While the 
compressibility $\kappa$ is related to the density variance, which in turn is 
affected by the average sign of each measurement, it is problematic to 
calculate $\kappa$ when opposite signs arise in the simulation. Therefore in 
our QMC calculations, instead of measuring the compressibility $\kappa$, we 
identify the boundary between the MI and NF phases by observing the saturation 
of the particle occupancy as temperature is reduced. In general, we restrict 
the maximum particle occupancy up to $6$ and set the lattice size $L=36$. 

\section{Finite-temperature phase diagrams}
\label{pd_den}
\begin{figure}
  \includegraphics[width=\linewidth]{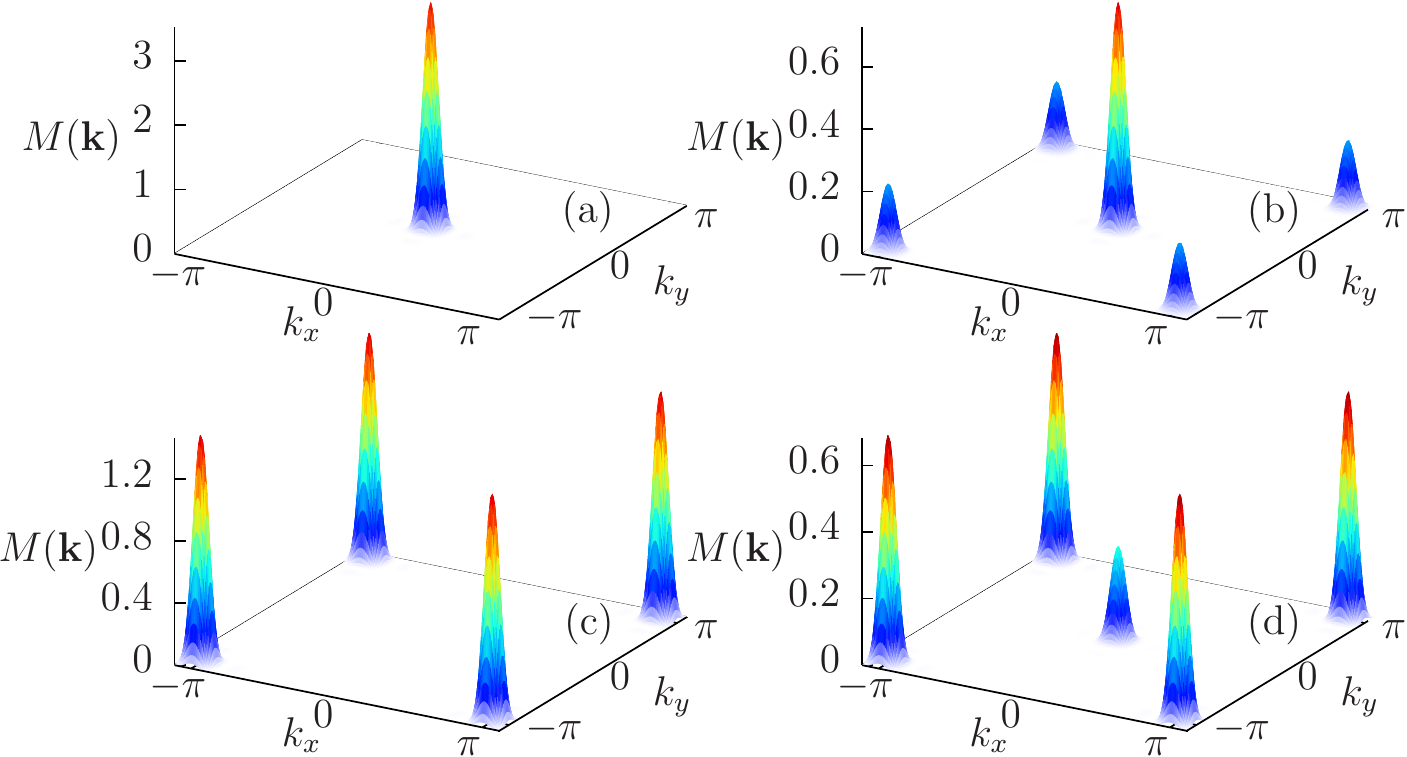}
  \caption{Examples of the momentum distributions of $M(\bf{k})$ for the 
	   compressible phases of dipolar bosons in two-dimensional optical 
	   lattices. (a) SF phase: the off-diagonal long-range order associated            
	   with superfluidity results in a sharp peak in $M(\bf{k})$ at the 
	   center of the Brillouin zone. (b) SS phase: the nearest-neighbor 
	   interaction leads to a supersolid compressible phase which exhibits 
	   a sharp peak in $M(\bf{k})$ at the center (due to the superfluid 
	   nature) and smaller peaks at the corners of the Brillouin zone (due 
	   to the long-range diagonal order). (c,d) SSF and SSS phases: in 
	   both cases the largest peaks move to the corners of the Brillouin 
	   zone.} 
  \label{k_space}
\end{figure}

  The competition between various terms in the model Hamiltonian leads to 
different phases at zero temperature. For $V=t'=0$, the competitions between 
the kinetic and on-site interaction energy leads to two quantum phases of 
BHM~\cite{fisher_89,jaksch_98}. Most of the phase diagram consists of 
Bose-Einstein condensed superfluid phase which possesses off-diagonal 
long-range order and is phase coherent. The momentum distribution of the 
single-particle correlation $M(\bf{k})$ for SF phase is shown in 
Fig.~\ref{k_space}(a) which shows a sharp peak of $M(\bf{k})$ at 
$\mathbf{k}=(0,0)$. This indicates a finite superfluid stiffness (or SF order 
parameter) $\rho_s\neq 0$ $(\phi\neq 0)$. For integer mean atomic densities and
strong repulsion, the Mott insulator phase appears in the lobe structures in 
$t-\mu$ plane of a homogeneous system. The size of the insulating lobes 
decreases with increasing $\mu$ as the corresponding larger mean-density favors 
superfluidity in the system. The MI phase is an incompressible phase with 
$M(\mathbf{k})=\rho_s=0$. At finite temperatures, the interplay of the quantum 
and thermal fluctuations results into a third, NF
phase~\cite{gerbier_07,sansone_08,ng_10,ng_11,mahmud_11}. This phase has zero 
$\rho_s$ even though the local atomic occupancy is incommensurate. In the 
finite-temperature phase diagrams, there is no broken symmetry between the 
insulating and NF phases, hence the transition across these phases exhibit a 
smooth crossover which is signaled by the change in the compressibility. 
Around the tip of the lobes, due to prevailing role of quantum fluctuations, 
the mean-field and quantum Monte Carlo approaches predict different 
phases~\cite{parny_12,luhmann_13}. This is due to the limitation of the 
mean-field theory, in which once the SF order parameter $\phi$ is zero, the 
hopping terms do not play any role. And, this underestimate the 
temperature-induced NF state around the tips of the insulating lobes.

The introduction of an off-site nearest neighbor interaction $V\neq0$ offers to 
stabilize quantum phases with spatial ordering, which is identified by the 
structure factor at finite momentum, for example for checkerboard compressible 
phases $S(\pi,\pi)\neq0$. This leads to two new phases, charge-density wave 
solid phase with integer or half-integer mean-densities, and supersolid phase 
which breaks two continuous symmetries: the phase invariance of the 
superfluidity and translational invariance to form crystalline 
order~\cite{lahaye_09,baranov_12,ng_08,bandyopadhyay_19,suthar_20a}. These two 
broken symmetries result into a sharp peak of  $M(\bf{k})$ at the center and 
four smaller peaks at finite $\mathbf{k}$ in two-dimensional Brillouin zone, as 
shown in Fig.~\ref{k_space}(b). Furthermore, the recent theoretical studies 
reported the presence of density-induced tunneling $(t'\neq0)$, at sufficiently 
higher densities leads to the existence of staggered superfluid (SSF) and 
staggered supersolid (SSS) phases~\cite{johnstone_19,kraus_20,suthar_20}. The 
emergence of staggered quantum phases is attributed to the destructive 
interference between single-particle hopping and DIT. The momentum 
distributions $M(\bf{k})$ of the  staggered phases reveal four sharp peaks at 
finite $\mathbf{k}$ and $M(\pi,\pi) > M(0,0)$ [Fig.\ref{k_space}(c,d)]. Here, 
we numerically determine the finite-temperature phase diagrams of isotropic 
dipolar bosons with correlated tunnelings. In particular, we choose higher 
average atomic density or chemical potential to examine the effects of DIT 
under the influence of thermal fluctuations. 
\begin{figure}
  \includegraphics[width=\linewidth]{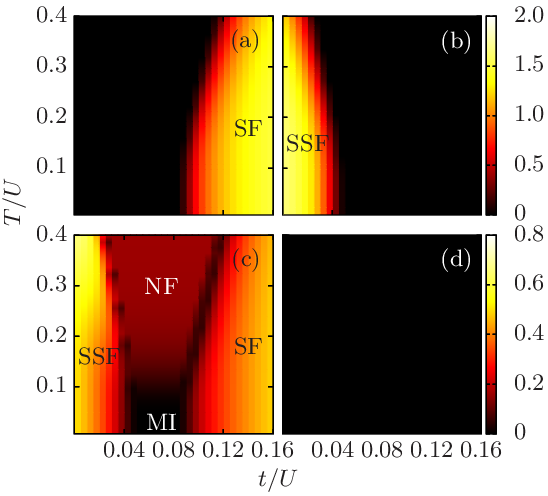}
  \caption{The thermal-averaged Fourier transform of the single-particle 
	   correlation  $M(\bf{k})$ at (a) $\mathbf{k} = (0,0)$ and (b) 
	   $\mathbf{k} = (\pi,\pi)$, (c) compressibility $\kappa$, and (d) the 
	   structure factor $S(\pi,\pi)$ as a function of hopping and 
	   temperature at weak NN interaction $V=0.24$. These are obtained 
	   using finite-temperature Gutzwiller mean-field theory in the 
	   grand-canonical ensemble. Here $\mu=3.5$ is considered. Different 
	   kind of superfluidity is separated by MI at lower $T$ and NF at 
	   higher $T$. Here, the maximum occupancy of bosons per site 
	   $n_{\rm max}=10$ is assumed. The system size is $L=36$.}
  \label{v_p24}
\end{figure}
\begin{figure}
  \includegraphics[width=\linewidth]{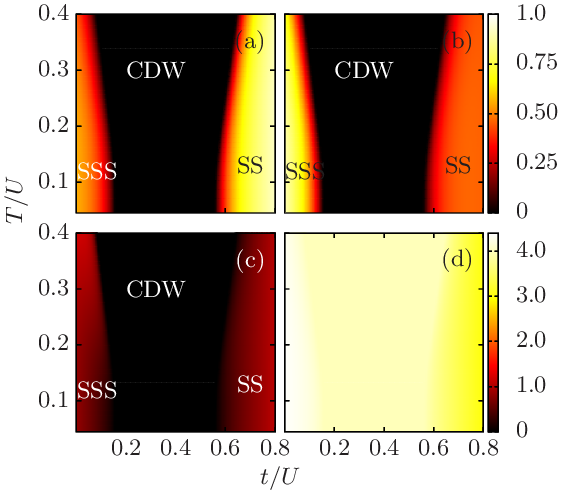}
  \caption{Shown here are various observables at strong NN interaction $(V=1)$ 
	   in hopping $(t)$ - temperature $(T)$ plane. The upper panel shows 
	   the Fourier transform of the single-particle correlation 
	   $M(\bf{k})$ at (a) $\mathbf{k}=(0,0)$ and (b) $\mathbf{k}=(\pi,\pi)$.           
	   In lower panel (c) the compressibility $\kappa$, and (d) structure 
	   factor at $\mathbf{k} = (\pi,\pi)$ are plotted. The chemical 
	   potential $\mu$ is chosen such that the effect of DIT introduces 
	   staggered phases. The normal and staggered supersolid phases are 
	   separated by incompressible CDW phase for the temperature limit 
	   considered. Here, the maximum occupancy per lattice site 
	   $n_{\rm max}=10$ and the system size $L=36$ are considered.}
  \label{v_1}
\end{figure}

In Fig.~\ref{v_p24}, we show the mean-field values of single-particle 
correlation $M(\bf{k})$, compressibility $\kappa$, and the structure factor 
$S(\pi ,\pi)$ as a function of hopping and temperature. The truncation in the 
Fock space $n_{\rm max} = 10$ is chosen such that the phases reported in the 
present work do not depend on it. In our calculations, $U$ sets the unit of 
energy scale, $U=1$, and periodic boundary conditions are assumed. We first 
consider $\mu=3.5$, $V=0.24$ and $t'=-0.02$, where previous studies show the 
existence of the staggered quantum phases at $T=0$ in square 
lattices~\cite{suthar_20}. In $T=0$ limit, a MI(2) phase appears in between 
two topologically distinct superfluid states. There exists a SSF phase for 
lower $t$, i.e. $t\sim|t'|$. This is confirmed by $M(0,0)<M(\pi,\pi)$, finite 
compressibility, and zero density-density correlation [$S(\pi,\pi)=0$]. 
However, at larger $t$, single-particle hopping driven normal superfluidity
emerges. The thermal fluctuations melt both superfluid regions and widen the 
NF parameter space at higher temperatures, as expected. For lower hopping 
strengths, the value of $M(\pi,\pi)$ decreases with temperature which suggests 
a shrink in the novel staggered superfluidity regime [Fig.~\ref{v_p24}(b)]. We 
further present the finite-temperature phase diagram at higher off-site NN 
interaction, $V=1$. It is pertinent to note that the DIT scales with NN 
interaction, hence we considered here $t'=-0.1$. The quantitative variations 
in the single-particle correlation $M(\bf{k})$ at $\bf{k}$=(0, 0) 
[Fig.~\ref{v_1}(a)] and $(\pi,\pi)$ [Fig.~\ref{v_1}(b)] suggest that the SSS 
phase is present at lower $t$ and the normal SS phase at larger $t$. The two 
compressible phases are separated by a phase with zero correlations and 
compressibility [Fig.~\ref{v_1}(c)] but finite structure factor $S(\pi,\pi)=4$ 
[Fig.~\ref{v_1}(d)]. This intervening phase is identified as CDW(4,0), which is
also confirmed by the density contrast at two consecutive sites. As $T$ 
increases, the CDW is robust to the thermal fluctuations and remains stable up 
to the maximum temperature range considered. However, at higher temperatures 
$T>0.4$, we expect the incompressible CDW phase and compressible SS and SSS 
phases to melt into the normal fluid state. 
\begin{figure}
  \includegraphics[width=\linewidth]{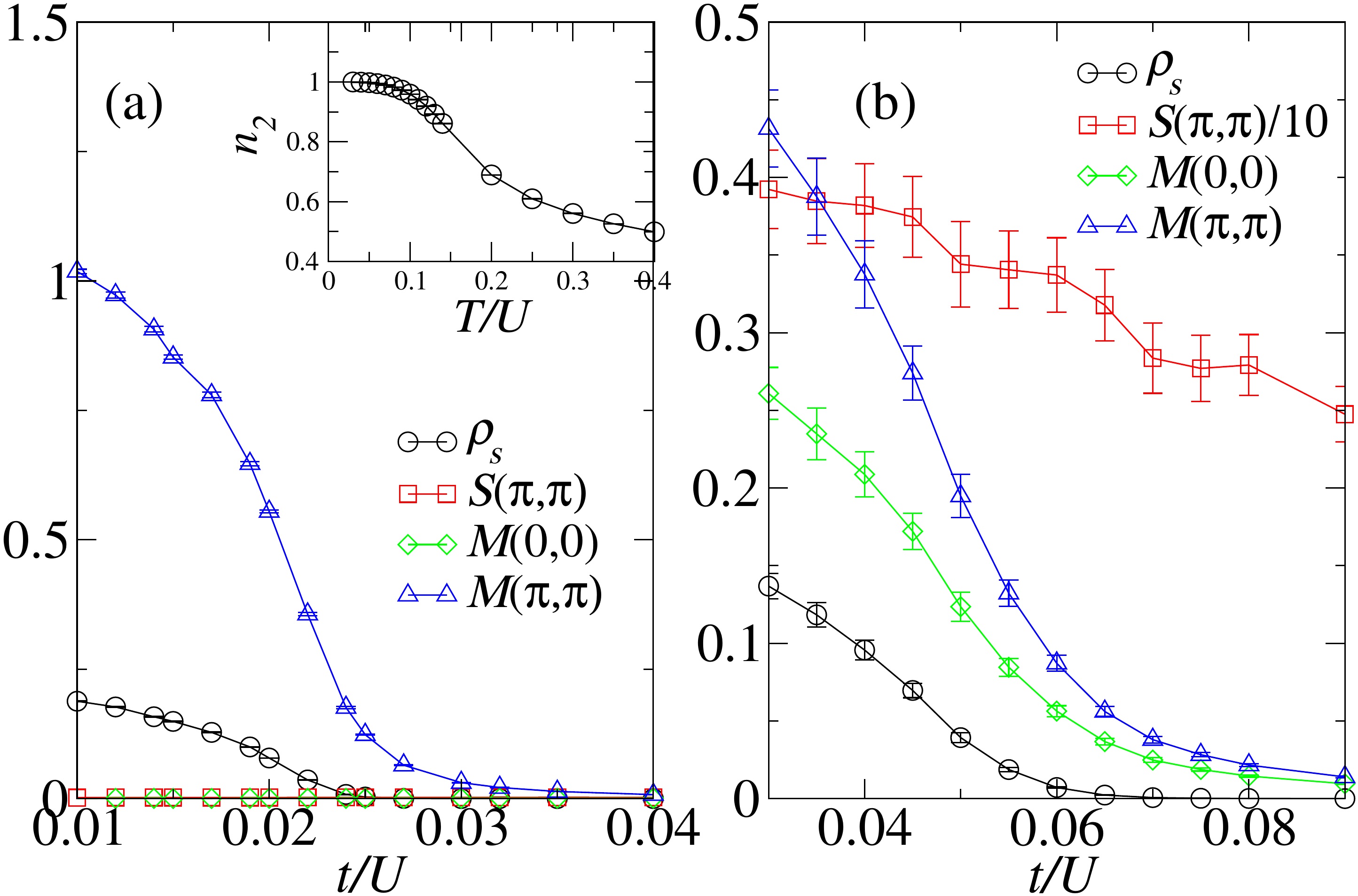}
  \caption{The order parameters obtained by QMC as a function of $t/U$ at 
	   $T/U=0.2$ for (a) $V=0.24$, $t'=-0.02$ and (b) $V=1$, $t'=-0.1$, 
	   showing the SSF-NF and SSS-CDW phase transitions, respectively. 
	   The inset in (a) shows the crossover from MI(2) to NF as the 
	   double-occupancy $n_2$ decreases from its saturation value.}
  \label{obs_T0.2}
\end{figure}

In Fig.~\ref{obs_T0.2} we present the QMC results showing the typical phase 
transitions of the SSF-NF (SSS-CDW) with $V=0.24$ ($V=1$) at finite temperature 
$T=0.2$. Consistent with the Fig.~\ref{v_p24} with small NN interaction 
$V=0.24$ [Fig.~\ref{obs_T0.2}(a)], single-particle correlation $M(\pi,\pi)$ is 
finite while $M(0,0)$ vanishes at small $t$, which signals the SSF phase as the 
DIT ($t'$) dominates the dynamics. When the domination of $t'$ ends at 
$t\sim 0.025$, the superfluid coherence is destroyed and the system becomes a 
NF. Similarly, in the case of strong NN interaction $V=1$ 
[Fig. ~\ref{obs_T0.2}(b)], the superfluidity order is destroyed as $t/U$ 
increases, except now the diagonal long-range order remains intact, which 
indicates a continuous phase transition from SSS to CDW(4,0), again in 
agreement with the mean-field results shown in Fig.~\ref{v_1}. Here,
the staggered character of the supersolid phase is verified by 
$M(0,0)<M(\pi,\pi)$.

\begin{figure}
  \includegraphics[width=\linewidth]{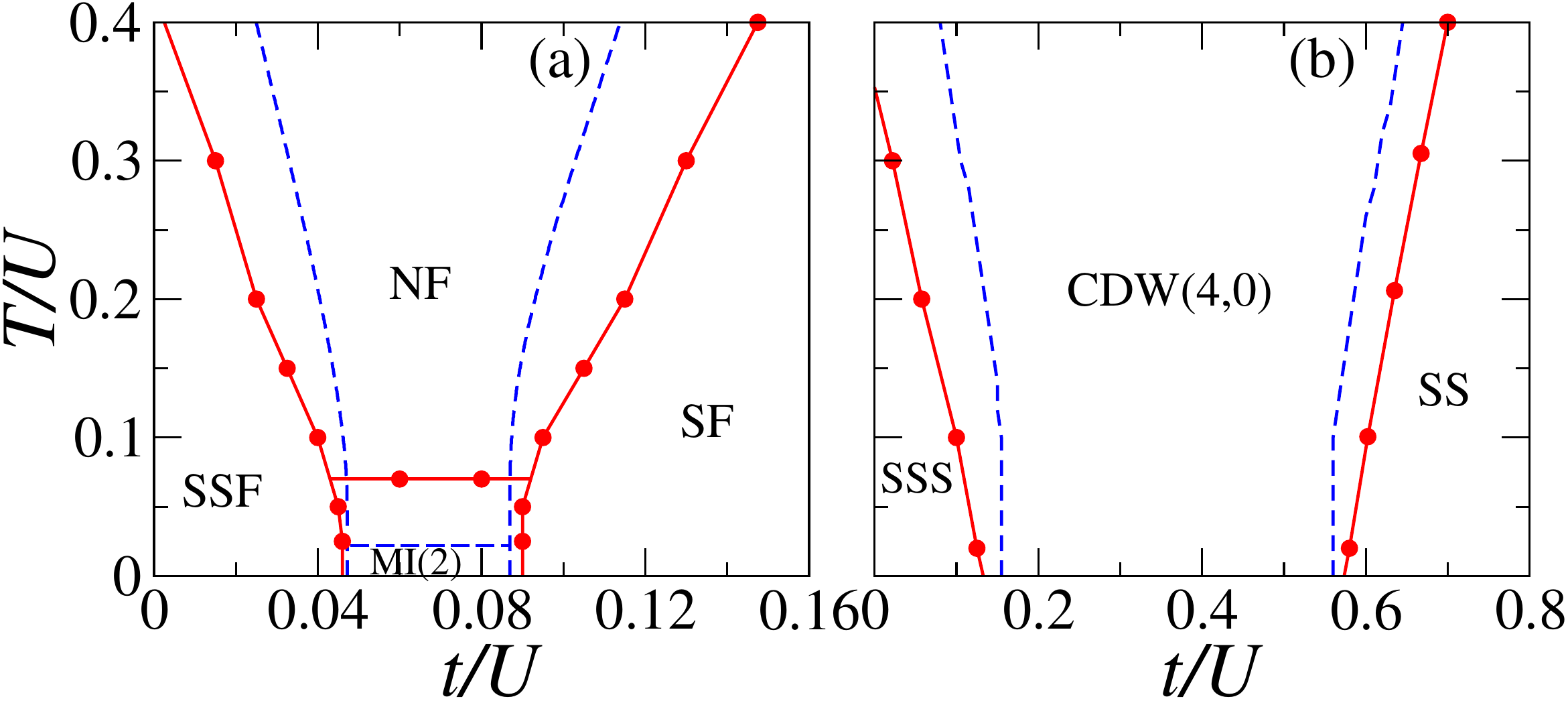}
  \caption{The finite temperature phase diagram at $\mu=3.5$ obtained using 
	   quantum Monte Carlo (solid lines) and Gutzwiller mean-field (dashed 
	   lines) approach. At finite temperatures, (a) for NN interaction 
	   comparable to the onsite interaction $(V=0.24)$, the compressible 
	   normal SF phase makes a transition to the SSF (as $t$ decreases) 
	   with an intervening NF phase. (b) When $V$ is strong $(V=1)$, a 
	   quantum phase transition between two types of supersolids occurs 
	   with an intervening CDW phase. The QMC corrects the prediction of the 
	   staggered phase regime using mean-field theory which overestimates the 
	   parameter space of compressible superfluid (supersolid) phases due to 
	   the poor resolution of atomic correlation between the lattice sites.
	   The phase boundaries are obtained for $L=36$.}
\label{pd_v_p24}
\end{figure}

Based on various observables using the mean-field and QMC simulations, we 
present the finite-temperature phase diagrams in Fig.~\ref{pd_v_p24}. In the 
phase diagram for weak NN interaction [Fig.~\ref{pd_v_p24}(a)], we find 
interesting features that the phase boundaries of the staggered superfluid and 
normal superfluid have opposite slopes on either side of the insulating (or NF) 
phase. It means when the single-particle hopping $t$ increases, the critical 
temperature of SSF reduces to zero at a critical $t$ and the system becomes a 
MI(2) solid. As $t$ increases until another critical $t$, the normal 
SF stabilizes with its critical temperature increases with $t$. This 
opposite dependence of critical temperature on both sides of insulating 
regime on the hopping $t$ can be understood by the competing role of $t$ and 
$t'$. As shown in Eq.~(\ref{effhop}), the effective hopping of bosons depends 
on the occupation of neighboring sites and the competing values of $t$ and $t'$. 
For the MI(2) phase, all sites are doubly occupied and the effective hopping 
vanishes if $t \sim -3t'$, i.e. near the middle of MI(2) phase with 
$t \sim 0.06$ in our case of weak NN interaction where $t'=-0.02$. For 
$t < 0.06$, the effective hopping becomes negative and results in SSF phase. 
The SSF phase, however, becomes less stable when $t$ increases as the quantum 
coherence deteriorates, resulting in a lower critical temperature. In contrast, 
when increasing $t > 0.06$ from the MI(2), the effective hopping is more 
positive and leads to a more stable SF and higher critical temperature. Therefore, 
the opposite dependence of the phase boundaries of SSF and SF on the hopping $t$ 
is a clear demonstration of the interplay of the hopping terms $t$ and $t'$. 
Similar mechanism also occurs for strong NN interaction [Fig.~\ref{pd_v_p24}(b)] 
where $t'=-0.1$ and the zero effective hopping appears at $t \sim 0.3$, again 
near the middle of the insulating CDW(4,0) phase. The only difference here is 
that the strong interaction leads to the diagonal long-range ordering in the 
whole phase diagram and the CDW survives even in high temperature regime in the 
diagram and hence NF phase is not observed. Nevertheless, using the mean-field 
approach we find that at high enough temperatures ($T/U\sim0.636$), thermal 
fluctuations will eventually destroy the insulating CDW phase and the system 
becomes NF.

Although both the result of mean-field and QMC simulation provide the same 
feature of phase boundaries as mentioned above,  there are quantitative 
differences between the two approaches. Previous studies have shown the 
importance of the QMC phase boundaries at finite 
temperatures~\cite{mahmud_11,parny_12,flottat_17}. The main difference can be
seen in Fig.~\ref{pd_v_p24} is that the mean-field overestimates the staggered 
and normal superfluid regions due to poor intersite correlations. For example, 
at $\mu=3.5$ the mean-field method shows SSF for $t=0.03$ and $T=0.3$, while 
the QMC predicts NF state. The critical hopping for the phase boundary 
separating SSF and NF obtained with QMC is lower and the deviations from the 
mean-field boundaries are prominent at higher temperatures. On the other hand, 
the critical value of $t$ for NF-SF transition is larger for QMC. It is 
important to note that the mean-field theory predicts larger SSS and SS domains 
due to absence of inter-site atomic correlations, however these predictions are 
corrected by QMC calculations, cf. Fig.~\ref{pd_v_p24}(b).

\section{Finite-size scaling analysis}
\label{fss_scal}
\begin{figure}
  \includegraphics[width=\linewidth]{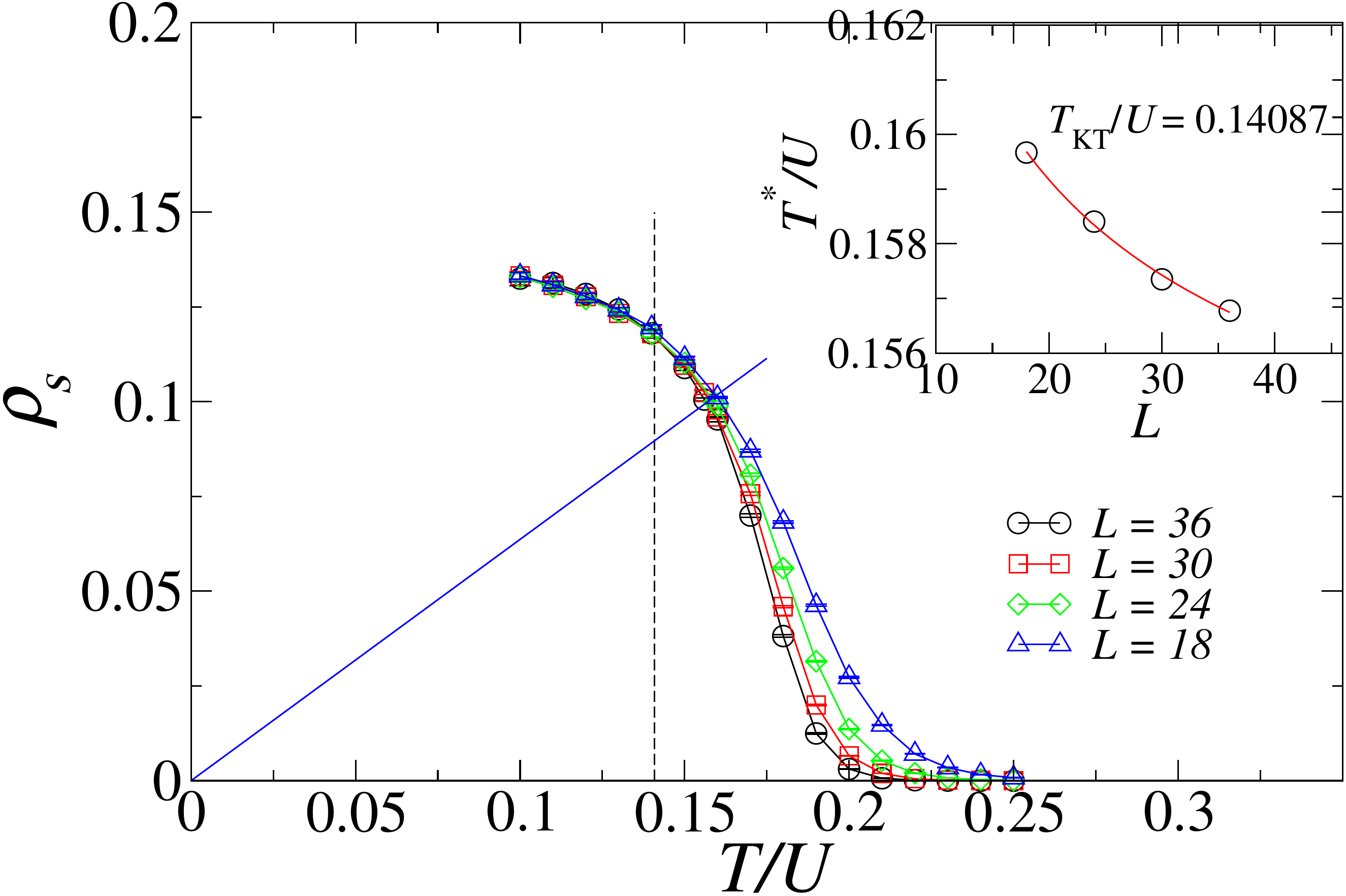}
  \caption{Finite-size scaling of the superfluidity $\rho_s$ as a function of 
	   $T$ for $V=0.24$, $t=0.025$ and $\mu=3.5$. The dashed line 
	   represents the fitted $T_{\text{KT}}$. The blue line is 
	   $\rho_s=\frac{2}{\pi}T$, and its intersection with the $\rho_s(T)$ 
	   for various $L$ gives $T^*(L)$, which shows excellent logarithmic 
	   dependence (inset) expected for KT transition (see the text).}
  \label{fss}
\end{figure}

A natural question arises about the thermal transition of staggered 
superfluid/supersolid to normal fluid/solid is whether it can be described by 
the Kosterlitz-Thouless transition as in the usual superfluid thermal phase 
transitions~\cite{PhysRevLett.88.167208}.  An instructive way to test the nature 
of the transition is via the finite-size scaling analysis, which we will 
describe here. In the thermodynamics limit, the superfluidity shows a universal 
jump at the KT transition temperature $T_{\text{KT}}$ that 
$\rho_s(T_{\text{KT}}) = 2T_{\text{KT}}/\pi$ which, however, is subjected to a 
logarithmic  correction 
$\rho_s(T_{\text{KT}},L) = \rho_s(T_{\text{KT}},\infty)\{1+1/[2\ln (L/L_0)]\}$ 
for finite size $L$ ~\cite{PhysRevB.37.5986,Hsieh_2013}. The constant value 
$L_0$ and the transition temperature $T_{\text{KT}}$ can be fitted by measuring 
the superfluidity of different system sizes $L$, as shown in Fig. \ref{fss}. The 
temperatures $T^*$ extracted from  the interaction points 
$\rho_s(T^*) = \frac{2}{\pi}T^{*}$ for various system sizes $L$ exhibit in a 
good agreement with the logarithmic correction 
$T^{*}(L) = T_{\text{KT}} \{1+1/[2\ln (L/L_0)]\}$ (inset) expected for KT 
transition. In the normal SF, it is clear that the thermal fluctuations lead to 
the unbinding of vortex and anti-vortex pairs via the KT 
transition~\cite{Kosterlitz_1973}. Our result suggests that the same mechanism 
is unaffected by shifting the condensation momentum from (0,0) to ($\pi,\pi$) 
such that the staggered ordering of the superfluidity does not alter the nature 
of the thermal phase transition.    

\section{Staggered phases in a trapping potential}
\label{trap}
\begin{figure}
  \includegraphics[width=\linewidth]{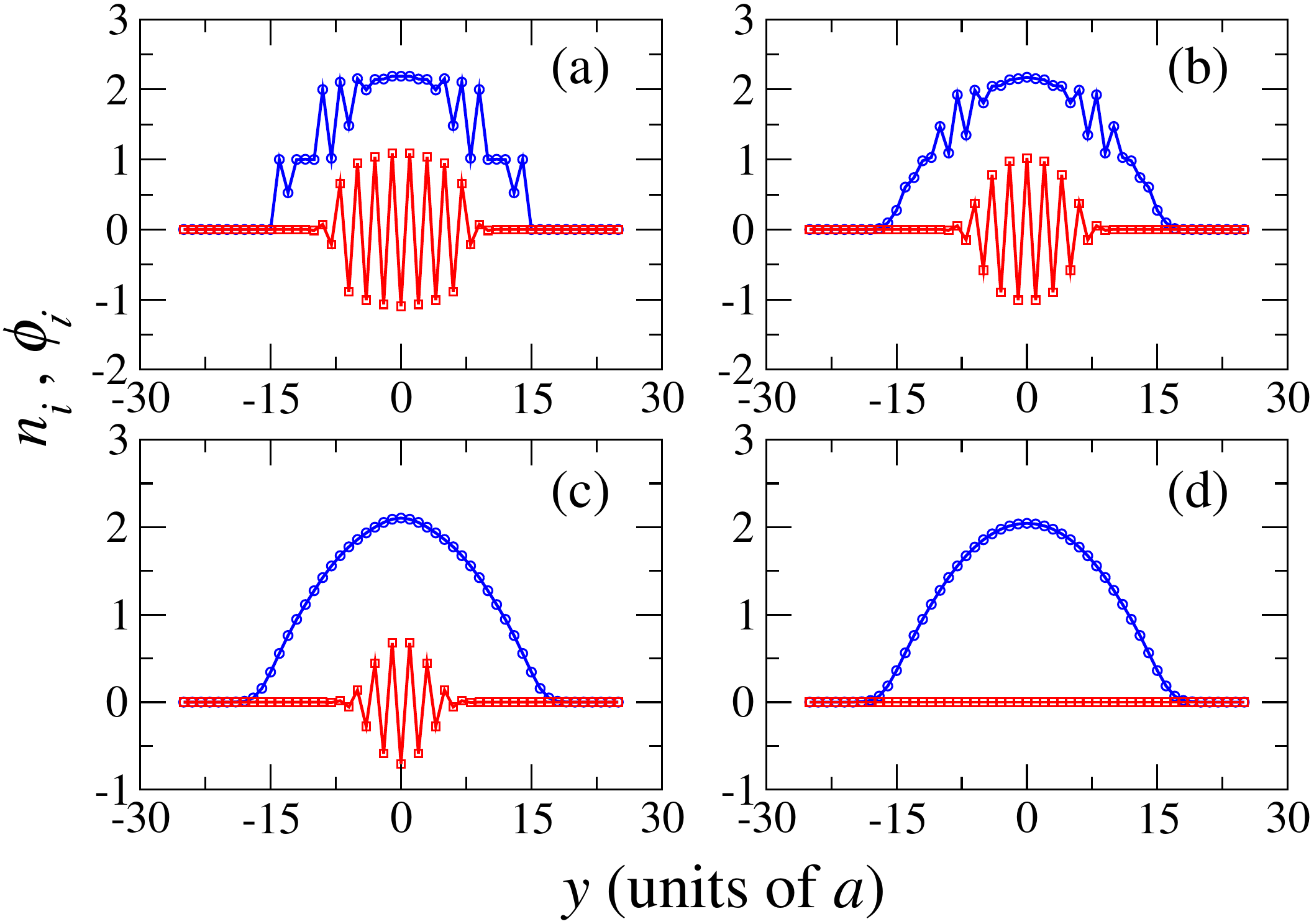}
  \caption{The average number occupancy and SF order parameter for $L=50$ square 
	   lattice along a one-dimensional cut at $x=0$ within a confining 
	   potential. The density-distribution (blue circle) and order parameter 
	   (red square) profiles are shown for different temperatures 
	   (a) $T=0.05$, (b) $T=0.2$, (c) $T=0.3$, (d) $T=0.35$. With an 
	   increase in $T$, the staggered variations of $\phi_i$ reduces, which 
	   indicates a staggered-to-normal phase transition. The staggered 
	   quantum phases vanishes due to thermal fluctuations at 
	   $T\approx 0.35$. The other parameters chosen are $t=0.025$, 
	   $\mu=3.5$, $V=0.24$, and temperatures are in units of $U$.
	   Here $y$ is in units of the lattice constant $a$.}
  \label{trap_ssf}
\end{figure}

We now consider the effects of realistic confinement on the staggered quantum 
phase transitions. The staggered phases result from the destructive 
interference between single-particle tunneling and DIT. In recent years, 
density-induced higher order processes have led to several novel phases, such 
as the interaction-driven mixing of orbitals~\cite{soltan_12} and correlated 
pair-tunneling leading to a twisted superfluid phase~\cite{jurgensen_12a}. 
Here, we discuss the parameter regime which can provide the staggered quantum 
phases in the presence of a magneto-optical trap in cold-atom experiments. We 
add a spatially varying chemical potential to the offset energy term of the 
model Hamiltonian~(\ref{ebhm}). The trapping potential is assumed as the 
harmonic potential $\epsilon_{i} = \Omega~i^2$, with $\Omega$ sets the strength 
of the potential.  In this external confinement the local chemical potential of 
the system changes as $\tilde{\mu} = \mu - \epsilon_{i}$. Co-existence of 
quantum phases occurs within the trapped system as a result of a change in 
$\tilde{\mu}$ within the trap. We have kept fixed the NN interaction $V=0.24$, 
$\mu=1$, and $\Omega=0.015$. The latter ensures that the trapped atomic density 
vanishes at the edge of the lattice. 

To illustrate the competition among different phases we have selected the cases 
$T=0.05,0.2,0.3$ and $0.35$. Using the mean-field approach, we compute the local 
atomic distributions and the SF order parameter in the trap. We plot the density 
and order parameter profiles for $L=50$ in Fig.~\ref{trap_ssf}. In particular, 
the plots are shown as a vertical cut at $x=0$ of the trapped square lattice, to 
observe the effects of the trapping potential. First we start considering 
$T=0.05$, where the observed effects are similar to the zero temperature case. 
From Fig.~\ref{trap_ssf}(a), we find a homogeneous atomic distribution $(n_i)$ at 
the center of the trap while the SF order parameter modulates between the same 
number and an alternating sign. These properties characterize the SSF at the 
center of the trap. In addition, the SSF phase is surrounded by a density 
modulation with a staggered order parameter distribution indicating the 
parameter domain of SSS. It is important to note that the SS phase also exhibits
modulations in $n_i$ and $\phi_i$ but the variation in $\phi_i$ between 
two real numbers does not change sign. In contrast, the staggered phase changes 
sign with the same (different) magnitude of the real numbers for SSF (SSS). 
Thus, the presence of harmonic potential exhibit the coexistence of two staggered 
quantum phases, SSF and SSS. At the edges of the trap, the superfluidity 
vanishes (as identified by $\phi_i=0$) and the insulating (MI and CDW) or normal 
state surrounds the staggered phases. 

We further show the effects of thermal fluctuations at finite temperatures on 
the stability of staggered phases. From Fig.~\ref{trap_ssf}(b) we observe the 
reduction in the regime with the modulation of $\phi_i$ at $T=0.2$, however the 
constant density at the center and the crystalline nature of the phase near the 
trap center still persist. It shows that the melting of staggered quantum 
phases begins from the edges, and it is more pronounced at $T=0.3$ 
[Fig.~\ref{trap_ssf}(c)]. At even higher temperatures, the staggered 
superfluidity vanishes and the normal fluid state occupies the trap due to the 
prevailing role of thermal fluctuations. The temperature corresponding to 
$T\approx0.3$ is in the few-nK regime for the extended Bose-Hubbard model with 
DIT for $^{168}${Er} atoms~\cite{baier_16}. Hence, the combined effects of the 
NN interaction as a genuine consequence of long-range interaction and 
interaction-driven DIT can lead to many-body staggered quantum phases in optical 
lattice experiments.

\section{Conclusion}
\label{conc}
We have studied the finite-temperature phase diagrams of soft-core dipolar 
bosons with density-induced tunneling in a square lattice potential. At weak 
dipolar interaction, the normal state intervenes between two topologically 
distinct superfluid states while at strong interaction the staggered and normal 
supersolid phases appear on either side of the insulating density-wave solid 
state. Both the mean-field calculation and quantum Monte Carlo simulation show 
that the critical temperature of the staggered superfluid phase decreases with 
single-particle tunneling while that of the normal superfluid increases, which 
is resulting from the interplay of DIT and single-particle tunneling. We have 
further shown, by using finite-size scaling of superfluid density, that the 
thermal phase transition of staggered superfluidity is KT-type, just like that 
of normal superfluid. This result suggests that the staggering of the 
superfluid phase does not alter the symmetry breaking process of the 
off-diagonal long-range order. Finally, we reveal the coexistence of quantum 
phases in the presence of an external trapping potential at finite 
temperatures. In quantum gas experiments, thermal fluctuations play a decisive 
role in the emergence of quantum phases. Recent experimental advances in 
the observation of density-induced tunneling and the novel superfluidity and 
phase transitions discussed in the present work may provide a way to realize 
staggered superfluids in ultracold dipolar experiments.

\begin{acknowledgements}
We thank M.-F. Yang and D. Angom for helpful discussions. K.S. acknowledges the 
support from IAMS, Academia Sinica and the Ministry of Science and 
Technology~(MOST), Taiwan, under the Grant No. MOST-109-2112-M-001-035-MY3. 
K.-K. N. acknowledges the support by the National Science and Technology Council 
under Grant no. 110-2112-M-029 -004 and 111-2112-M-029 -007. 
\end{acknowledgements}


\bibliography{dip_2d}{}

\end{document}